\documentclass[]{spie}  

\usepackage{amsmath,amsfonts,amssymb}
\usepackage{graphicx}
\usepackage[colorlinks=true, allcolors=blue]{hyperref}

\title{Final design and development status of the acquisition and guiding system for SOXS.}

\author[*,a,b]{Anna~Brucalassi}
\author[a,c]{Giuliano~Pignata}
\author[c,v]{José~Antonio~Araiza-Duran}

\author[d]{Sergio~Campana}
\author[e]{Riccardo~Claudi}
\author[f]{Pietro~Schipani}
\author[d]{Matteo~Aliverti}
\author[e]{Andrea~Baruffolo}
\author[g]{Sagi~Ben-Ami}
\author[h]{Federico~Biondi}
\author[f]{Giulio~Capasso}
\author[f]{Mirko~Colapietro}
\author[i]{Rosario~Cosentino}
\author[j]{Francesco~D'Alessio}
\author[d]{Paolo~D'Avanzo}
\author[d]{Matteo~Genoni}
\author[k]{Ofir	Hershko}
\author[l,m]{Hanindyo~Kuncarayakti}
\author[d]{Marco~Landoni}
\author[n]{Matteo~Munari}
\author[e]{Kalian Radhakrishnan}
\author[k]{Michael~Rappaport}
\author[e]{Davide~Ricci}
\author[o]{Adam~Rubin}
\author[p]{Salvatore~Scuderi}
\author[j]{Fabrizio~Vitali}
\author[n]{Ricardo~Zanmar~Sanchez}
\author[q]{David~Young}
\author[r]{Jani~Achrén}
\author[s]{Iair~Arcavi}
\author[k]{Rachel~Bruch}
\author[e]{Enrico~Cappellaro}
\author[f]{Massimo~Della~Valle}
\author[e]{Marco~De~Pascale}
\author[n]{Rosario~Di~Benedetto}
\author[f]{Sergio~D'Orsi}
\author[k]{Avishay~Gal-Yam}
\author[i]{Marcos~Hernandez}
\author[l,m]{Jari~Kotilainen}
\author[t]{Gianluca~Li~Causi}
\author[m]{Seppo~Mattila}
\author[d]{Marco~Riva}
\author[e]{Bernardo~Salasnich}
\author[q]{Stephen~Smartt}
\author[u]{Maximilian~Stritzinger}
\author[i]{Hector~Ventura}

\affil[a]{Universidad Andres Bello, Avda. Republica 252, Santiago, Chile }
\affil[b]{INAF -- Osservatorio Astronomico di Arcetri, Largo Fermi 5, I-50125, Florence, Italy }
\affil[c]{Millennium Institute of Astrophysics (MAS)}
\affil[d]{INAF -- Osservatorio Astronomico di Brera, Via Bianchi 46, I-23807, Merate, Italy }
\affil[e]{INAF -- Osservatorio Astronomico di Padova, Vicolo dell’Osservatorio 5, I-35122, Padua, Italy }
\affil[f]{INAF -- Osservatorio Astronomico di Capodimonte, Sal. Moiariello 16, I-80131, Naples, Italy }
\affil[g]{Harvard-Smithsonian Center for Astrophysics, Cambridge, USA }
\affil[h]{Max-Planck-Institut f\"ur Extraterrestrische Physik, Giessenbachstr. 1, D-85748 Garching, Germany}
\affil[i]{FGG-INAF, TNG, Rambla J.A. Fernández Pérez 7, E-38712 Breña Baja (TF), Spain }
\affil[j]{INAF -- Osservatorio Astronomico di Roma, Via Frascati 33, I-00078 M. Porzio Catone, Italy }
\affil[k]{Weizmann Institute of Science, Herzl St 234, Rehovot, 7610001, Israel }
\affil[l]{Finnish Centre for Astronomy with ESO (FINCA), FI-20014 University of Turku, Finland}
\affil[m]{Tuorla Observatory, Dept. of Physics and Astronomy, FI-20014 University of Turku, Finland }
\affil[n]{INAF -- Osservatorio Astrofisico di Catania, Via S. Sofia 78 30, I-95123 Catania, Italy }
\affil[o]{European Southern Observatory, Karl Schwarzschild Str. 2, D-85748, Garching bei M\"unchen}
\affil[p]{INAF - Istituto di Astrofisica Spaziale e Fisica Cosmica,Via Corti 12, I-20133 Milan, Italy}
\affil[q]{Queen's University Belfast, Belfast, County Antrim, BT7 1NN, UK}
\affil[r]{Incident Angle Oy, Capsiankatu 4 A 29, FI-20320 Turku, Finland }
\affil[s]{Tel Aviv University, Department of Astrophysics, 69978 Tel Aviv, Israel }
\affil[t]{INAF - Istituto di Astrofisica e Planetologia Spaziali, Rome, Italy}
\affil[u]{Aarhus University, Ny Munkegade 120, D-8000 Aarhus, Denmark }
\affil[v]{Centro de Investigaciones en Optica A. C., Loma del Bosque 115, Lomas del Campestre, 37150 Leon Guanajuato, Mexico}

\authorinfo{$^*$Contact information:
  A.B: anna.brucalassi@gmail.com, 
  Telephone: +39 055 2752 243\\
  G.P: pignata@gmail.com, 
  Telephone: +33 (0)1 98 76 54 32
 }

\begin{document} 
\maketitle

\begin{abstract}
SOXS (Son Of X-Shooter) will be the new medium resolution (R$\sim$4500 for $1''$ slit), high-efficiency, wide band spectrograph for the ESO NTT at La Silla, optimized for classification and follow-up of transient events. SOXS will simultaneously cover UV-optical and NIR bands (0.35-2.00 micron) using two different arms and a pre-slit Common Path feeding system. The instrument will be also equipped by a Calibration Unit and an Acquisition Camera (AC) System.
In this paper we present the final opto-mechanical design for the AC System and we describe its development status. 
The project is  currently  in  manufacturing  and  integration  phases.  

\end{abstract}

\keywords{SOXS, Spectroscopy, Imaging, Acquisition and Guiding}

\section{INTRODUCTION}
\label{sec:intro}  

The \emph{ Son Of X-Shooter} (SOXS), is the new instrument\cite{Schipani2018,2016Schipani}, actually under manufacturing  and  integration  phase, for the European Southern Observatory (ESO) New Technologies Telescope (NTT) at the La Silla Observatory, Chile. It will be dedicated to the follow up of any kind of transient events ensuring fast response time, high efficiency and availability. 
SOXS will simultaneously cover the electromagnetic spectrum from 0.35 to 2.0$~\mu$m with a spectral resolution of R$\sim$4500, using two arms (UV-VIS and NIR)\cite{Rubin2018,Sanchez2018} and a pre-slit Common Path (CP) feeding system\cite{Claudi2018,Biondi2018}.

SOXS is also equipped with a Calibration sub-unit and an Acquisition Camera system (CAM)\cite{Brucalassi2018} attached directly to the CP. The CAM system is foreseen to work not only for target acquisition and (optional) secondary guiding, but it will be also equipped with a filter wheel and a scientific camera for providing photometric observations.

The schedule for the development of the CAM system has been seriously affected by the ongoing COVID19 pandemic, interrupting some providers activities and resulting in a delay of about some months.
In the following we summarize the final opto-mechanical design for the AC System and we describe its development status. 
\begin{figure} [ht]
   \begin{center}
   \begin{tabular}{c} 
   \includegraphics[height=7cm]{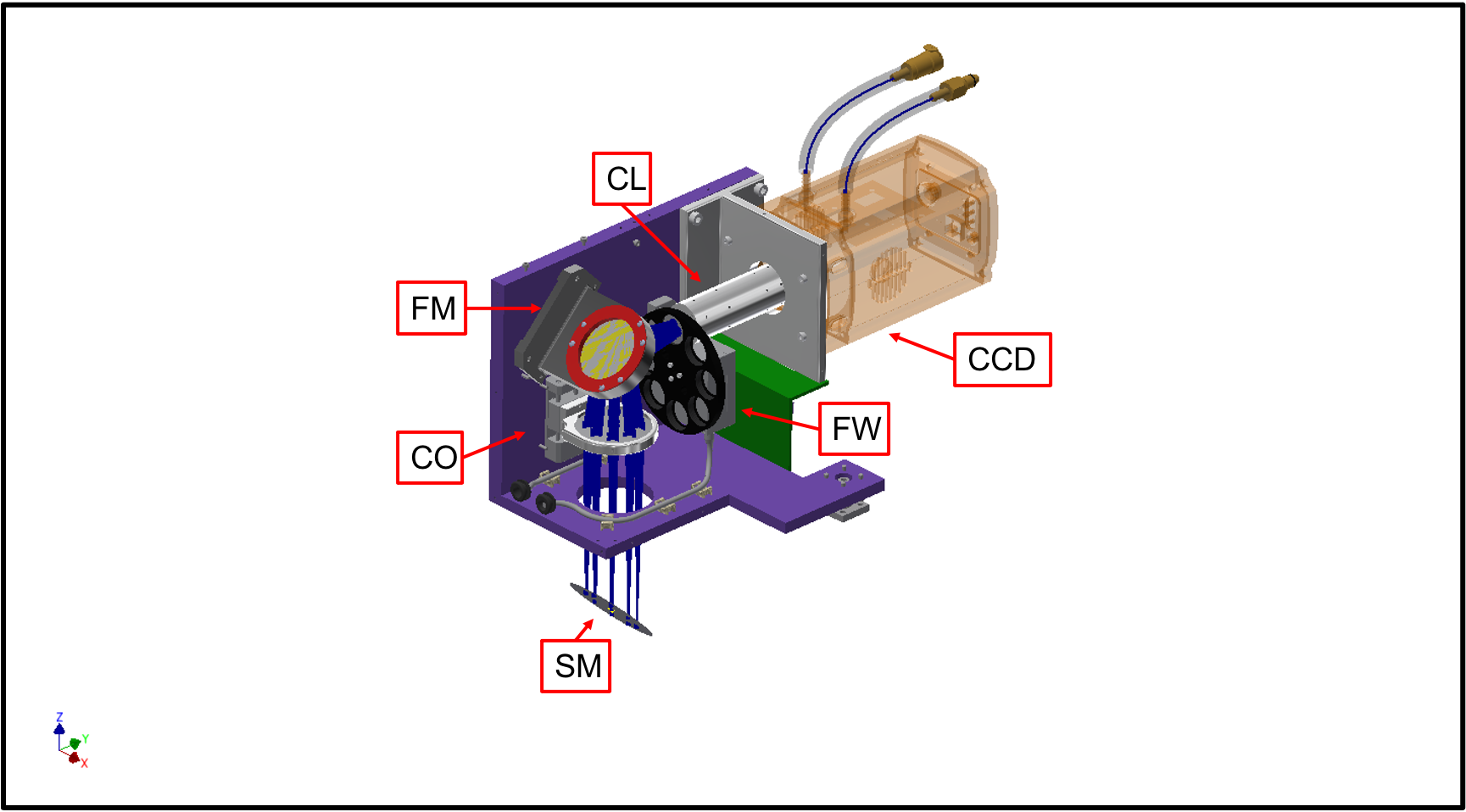}
   \end{tabular}
   \end{center}
   \caption[example] 
   { \label{fig:ASM_CAM_3.5amin_SOXS2} 
Overview of the Acquisition Camera System:  it consists of a collimator lens (CO), folding mirror (FM), filter wheel (FW), focal reducer optics (CL) and CCD camera.  
All the elements are included in a structure made of 6061-T6 Aluminum.}
   \end{figure} 

\section{Overview}
\
Figure~\ref{fig:ASM_CAM_3.5amin_SOXS2} shows an overview of the Acquisition Camera System. 
It is based on a collimator lens (CO), a folding mirror (FM), a filter wheel (FW)  with a broad-band filter set (ugrizY and V-Johnson), a focal reducer optics (CL) and CCD camera.  
All the system is included in a structure made of 6061-T6 Aluminum. The detector is a 1k x 1k Andor iKon-M 934 camera\footnote{https://andor.oxinst.com} and is already in house. The driver to run the camera under the SOXS control software has been developed and validated\cite{Capasso2018,Ricci2018}.

A system named as \emph{CAM selector} and placed in the CP at the level of of the Nasmyth focal plane will redirect the F/11 beam from the telescope toward the Acquisition and Camera system.

\begin{figure} [ht]
   \begin{center}
   \begin{tabular}{c} 
    \includegraphics[height=4.8cm]{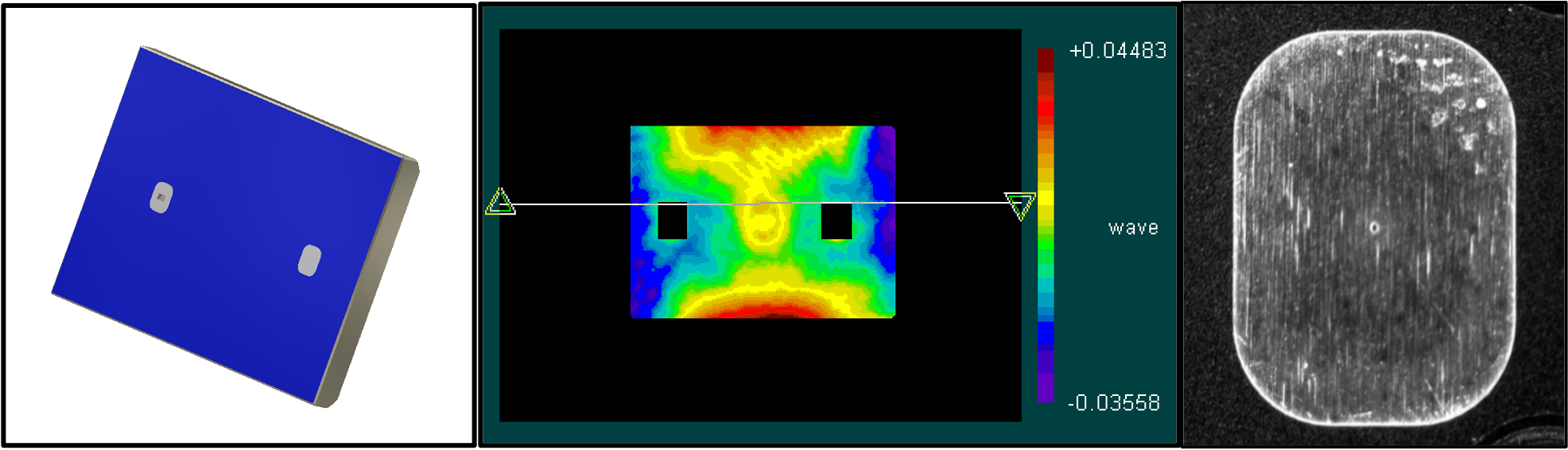}
    
   \end{tabular}
   \end{center}
   \caption[CAMSelectorMirror] 
   { \label{fig:CAM_SelectorMirror} 
 Left panel: view of the selector mirror. Central pannel: Wavefront Map of the selector mirror: PV = 0.161 fr, rms = 0.014 wave . Right panel: zoomed view of the metallic plate for the \emph{Artificial Star Mode}. }
   \end{figure} 

The CAM selector system consists of a linear stage that
carries a pellicle beam-splitter and 
a single mirror with three positions for different functions (Acquisition and Imaging, Artificial Star and Spectroscopy). 
The mirror and the pellicle beam-splitter are tilted at 45$^\circ$ and direct light from sky or from the slits, respectively, to the CAM optics.
The pellicle beam-splitter allows us to use the CAM system as slit viewing camera (with the calibration lamp on).
In \emph{Acquisition and Imaging Mode}, the selector mirror redirects the full field towards the CCD and the CAM system can be operated for photometric observations. 
With the selecor mirror in \emph{Artificial Star Mode} a 0.5 arcsec reference pinhole acts as an artificial star on the focal plane by switching on the calibration lamp.
Finally, in  \emph{Spectroscopy  Mode},  the  selector  mirror  passes  an  unvignetted  field  of  15  arcsec  to  the spectrograph  slits,  whereas  the  peripheral  field  is  simultaneously  imaged  on  the  camera.   Thus, off-axis  secondary guiding is possible using peripheral sources.
We refer to~\citenum{Brucalassi2018,Aliverti2018} for a more detailed description of the CAM selector system and its functionalities.

The selector mirror has been already manufactured and is based on a rectangular Fused Silica mirror (91.0x70.0x10.0mm) with two holes that allow the optical beam to pass. On the reflective surface two specific engravings ensure two corresponding metallic plates to be fixed by glue. On the center of the metallic plates a 0.5 arcsec reference pinhole and a rectangular slit-hole for a 15 arsec unvignetted beam have been created by laser cut.
Figure~\ref{fig:CAM_SelectorMirror} shows a view of the selector mirror (left panel), its Wavefront Map (central panel) and a zoomed picture of the metallic plate for the \emph{Artificial Star Mode} (right panel).

\section{Final Optical Design}

After the Final Design Review, the CAM optical design needed some modifications due to the unfeasible waiting time for the procurement of some glasses. However, the general architecture and concept of the design has been preserved.
\begin{figure} [ht]
   \begin{center}
   \begin{tabular}{c} 
   \includegraphics[height=7.0cm]{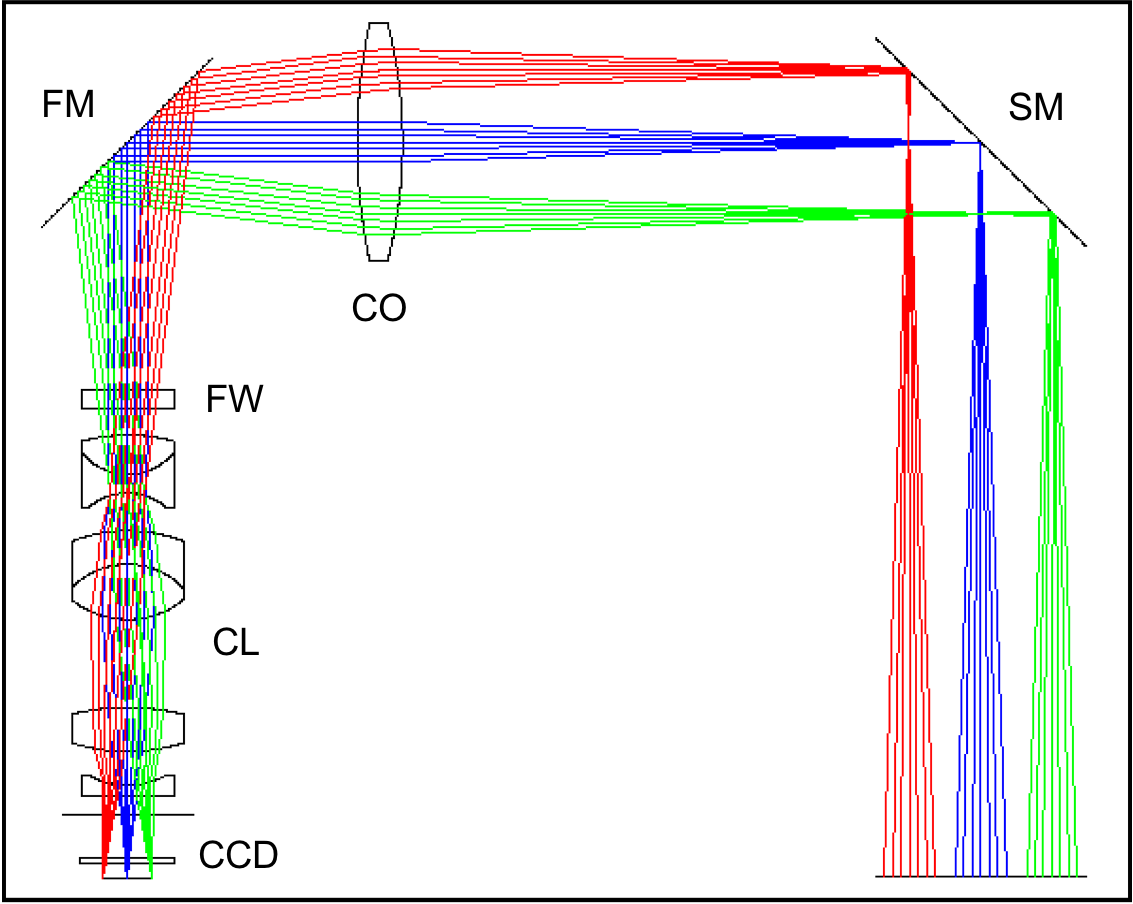}
   \end{tabular}
   \end{center}
   \caption[example] 
   { \label{fig:CAM_OpticalDesign} 
Optical Layout of the CAM System. From right to the left: the Selector Mirror (SM), the Collimator lens (CO), the Folding Mirror (FM), the Filter Wheel (FW), the Camera Lens (CL) with two doublets (D1, D2), the two singlets (S1, S2), the detector window and the focal plane.}
   \end{figure} 
The new optical design is illustrated in Figure~\ref{fig:CAM_OpticalDesign}  and in Table~\ref{tab:OpticalElements_data}.
\begin{table}[ht]
\caption{Optical Elements data.} 
\label{tab:OpticalElements_data}
\begin{center}       
\begin{tabular}{|l|l|l|l|l|} 
\hline
\rule[-1ex]{0pt}{3.5ex}  Description & Radius(mm) & Distance (mm) & Material & Diameter(mm)\\
\hline
\rule[-1ex]{0pt}{3.5ex}  Telescope Focus & - & 155.81 & - & -   \\
\hline
\rule[-1ex]{0pt}{3.5ex} Collimator & 132.10 & 12.24 & N-FK58 & 65.00  \\
\hline
\rule[-1ex]{0pt}{3.5ex}       & -158.48 & 61.84 &   -  & 65.00 \\
\hline
\rule[-1ex]{0pt}{3.5ex} Folding Mirror & - & -67.37 & -& 65.00 \\
\hline
\rule[-1ex]{0pt}{3.5ex} Filter  & - & -5.00 & N-BK7  & 22.00 \\
\hline
\rule[-1ex]{0pt}{3.5ex}         &   & -7.38 &  -     & 22.00  \\
\hline
\rule[-1ex]{0pt}{3.5ex}    D1   & -44.91  & -10.74 &  N-FK58 & 25.00 \\
\hline
\rule[-1ex]{0pt}{3.5ex}         & 16.01  & -5.09 & K10 & 25.00 \\
\hline
\rule[-1ex]{0pt}{3.5ex}         & -15.97  & -10.27 & - & 21.00 \\
\hline
\rule[-1ex]{0pt}{3.5ex}    D2   & -40.98  & -9.11 & K10 & 30.00\\
\hline
\rule[-1ex]{0pt}{3.5ex}         & -20.54  & -15.02 & N-FK58 & 30.00\\
\hline
\rule[-1ex]{0pt}{3.5ex}         & 23.98 & -24.00 & - & 30.00 \\
\hline
\rule[-1ex]{0pt}{3.5ex}    S1   & -62.09& -12.00 &LLF1 & 30.00 \\
\hline
\rule[-1ex]{0pt}{3.5ex}         & 48.73  & -9.19 & - & 30.00 \\
\hline
\rule[-1ex]{0pt}{3.5ex}    S2   & 21.25  & -2.97 & PBL6Y& 21.00\\
\hline
\rule[-1ex]{0pt}{3.5ex}         & - & -4.93 & - & 25.00 \\
\hline
\rule[-1ex]{0pt}{3.5ex} Shutter & - & -11.85 & -& -  \\
\hline
\rule[-1ex]{0pt}{3.5ex} Detector Window   &  - & -1.50 & SILICA       & 25.40 \\
\hline
\rule[-1ex]{0pt}{3.5ex}         & - & -4.15 & - &  25.40\\
\hline
\rule[-1ex]{0pt}{3.5ex}  Focal Plane & - &  &   & - \\
\hline

\end{tabular}
\end{center}
\end{table}

The first element after the CAM selector mirror is a collimator lens of 65mm diameter that reimages the pupil onto a compact camera. A folding mirror and a filter wheel, illustrated in Figure~\ref{fig:CAM_OpticalDesign} by a single plane glass, follow in the optical path.
The subsequent camera lens, relays the Nasmyth focus on the detector, with an $F_{\#}$=3.6, through 2 doublets and two singlets of max diameter 30.0 mm.
A Field of View of 3.5x3.5 arcmin (linear) can be reached in imaging mode, resulting in a pixel scale of $\sim$0.2 arcsec.\\ The total length from focus to focus is 430.46 mm. 

\subsection{Image Quality and Thermal Analysis}
\label{sec:CAM_ImageQuality}

\begin{figure} [ht]
   \begin{center}
   \begin{tabular}{c} 
   \includegraphics[height=6.2cm]{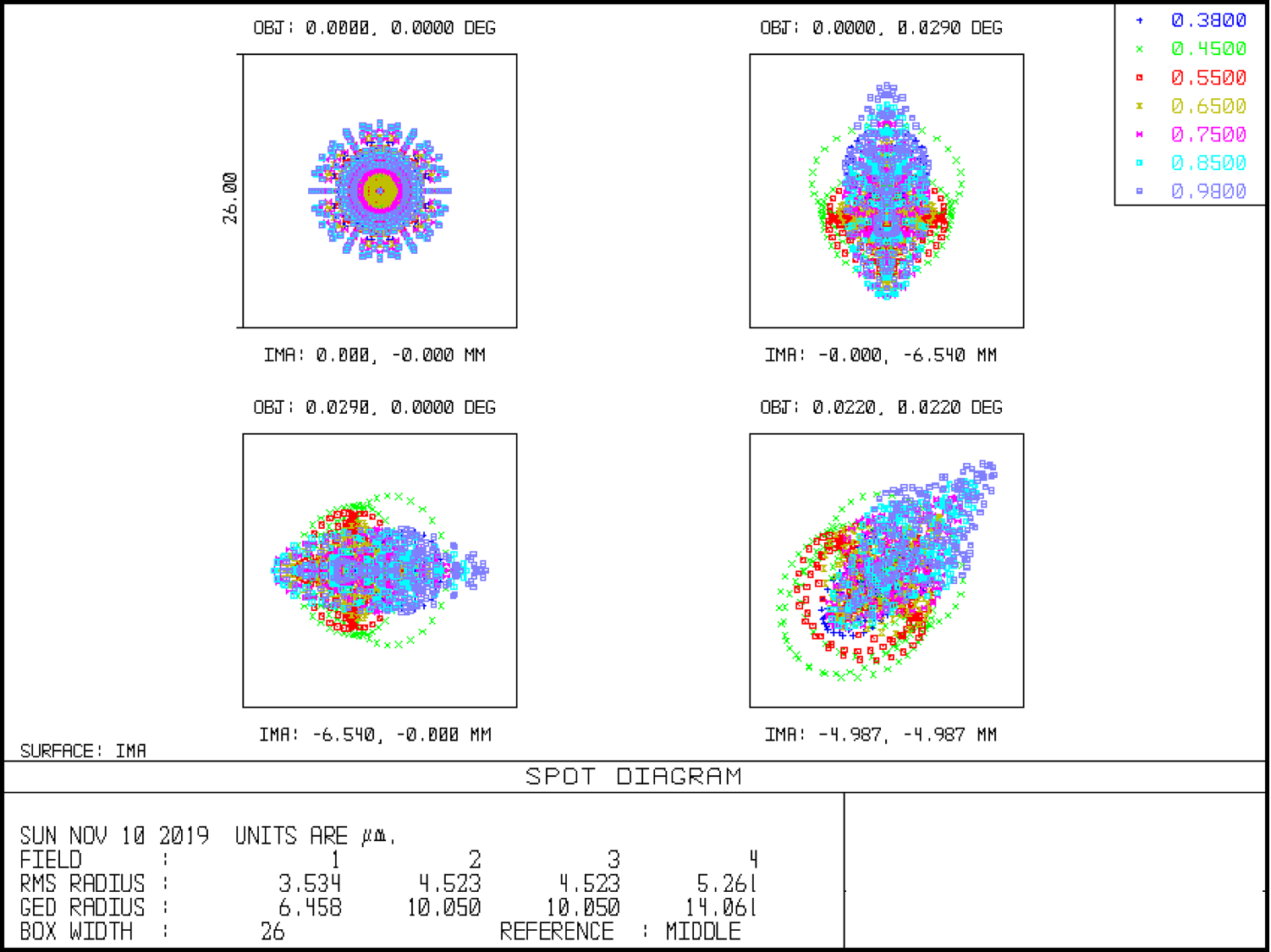}
   \includegraphics[height=6.2cm]{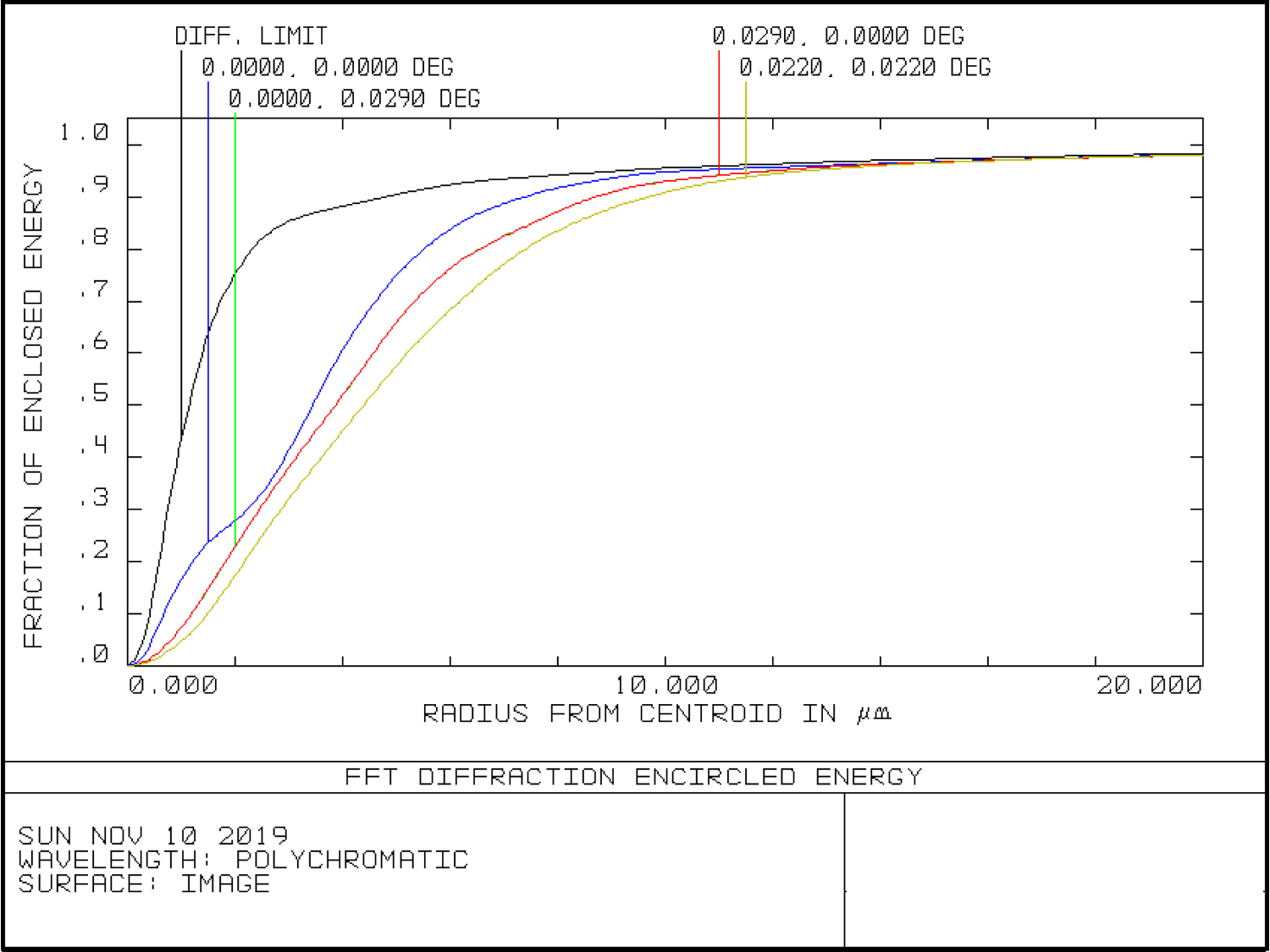}
   \end{tabular}
   \end{center}
   \caption[CAMSpotDiagram] 
   { \label{fig:CAM_SpotDiagram} 
 Left panel: CAM Spot diagrams for field positions correspond to on-axis, 3.5 arcmin (side of the detector) in x and y direction and along the diagonal corresponding to a 3.75 arcmin diameter (corner of the detector).
Right panel: Geometric encircled  energy.}
   \end{figure}

Figure~\ref{fig:CAM_SpotDiagram} shows the spot diagrams and the geometric encircled energy for the new optical configuration. The design has been done considering 10$^\circ$C and 0.76 atm.
Plotted field positions correspond to the center of the field, sides of the detector in x and y at 3.5 arcmin and the position along the diagonal corresponding to a 3.75 arcmin diameter, for which the quality is still good.
Clearly, the spot is contained in a box of 2x2 pixels (26.0 $\mu$m) for all the analyzed views. 

\begin{figure} [ht]
   \begin{center}
   \begin{tabular}{c} 
   \includegraphics[height=6.2cm]{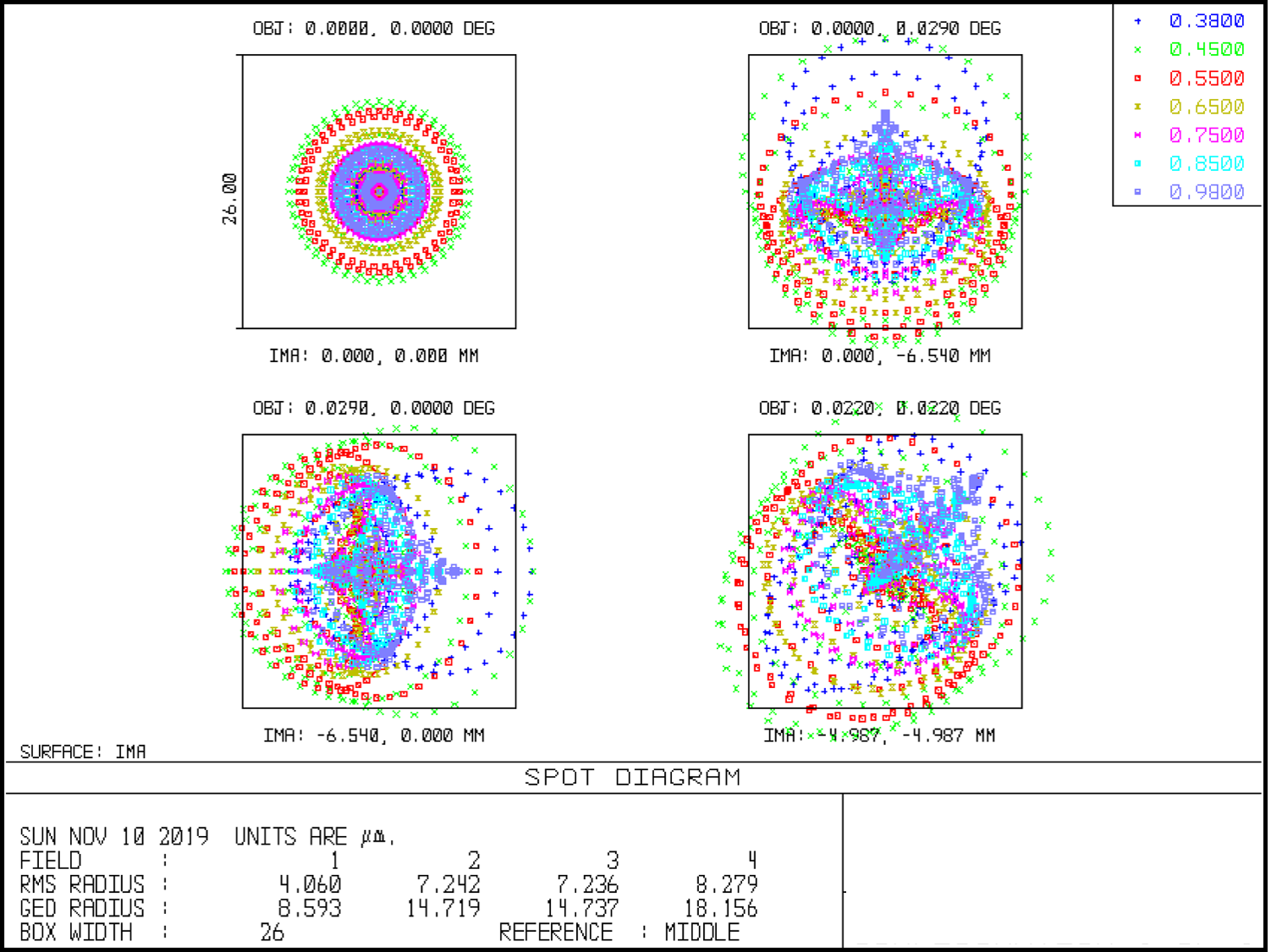}
   \includegraphics[height=6.2cm]{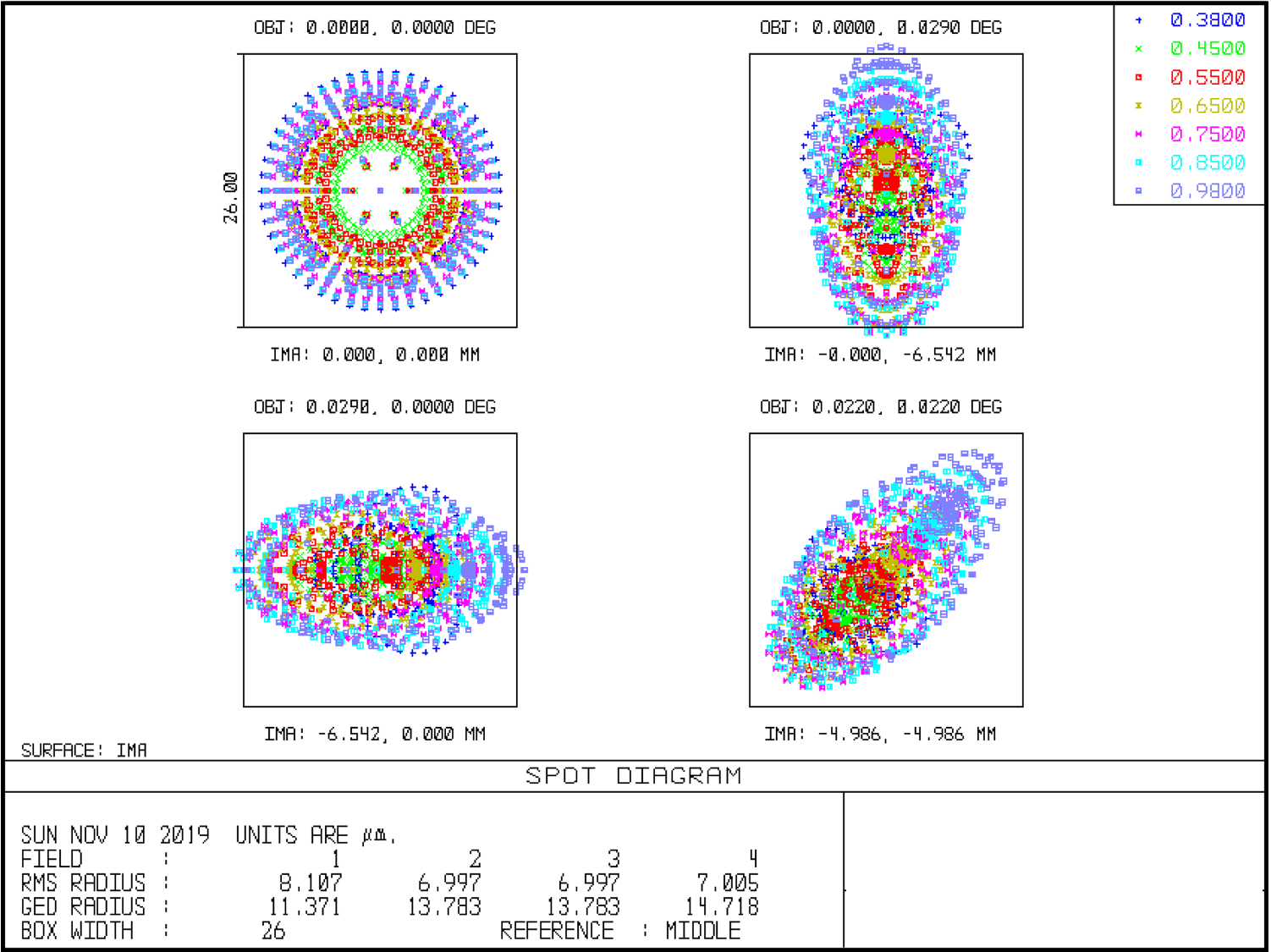}
   \end{tabular}
   \end{center}
   \caption[CAMThermalAnalysis] 
   { \label{fig:CAM_TempChanges} 
Spot diagrams considering temperatures of 0$^\circ$ C (left panel) and 20$^\circ$ C (right panel). The geometrical dimension of the spot are still about 2 pixels (26.0$\mu$m). }
   \end{figure}

 Figure~\ref{fig:CAM_TempChanges} instead presents the spot diagrams of the new optical configuration considering temperatures of 0$^\circ$ C (left panel) and 20$^\circ$ C (right panel).
 A change of $\pm10^\circ$C causes a deterioration of the image, however, always with a geometrical dimension of about 2 pixels (26.0 $\mu$m). In addition, the optical quality can be completely recovered re-focusing with the collimator lens.  

\section{Development Status}
\label{sec:Mec_Layout}

At the time of writing, all the optical parts together with their opto-mechanical mounts are in production phase. 
The external structure that will host the CAM system will be made by Aluminum 6061-T6 with a T-Shape as shown in Figure~\ref{fig:ASM_CAM_3.5amin_SOXS2}. All the walls are structural with thickness of $\sim$15 mm, while the cover is not structural and is made of a 3 mm thick Aluminum plate.

The detector will be mounted on a support directly connected to the external structure.
The centering of the CCD will be done shimming the 3 holes and 3 dowel pins used to place the CCD together with the support system.

\subsection{Collimator Lens}
\begin{figure} [h]
   \begin{center}
   \begin{tabular}{c} 
   \includegraphics[height=5cm]{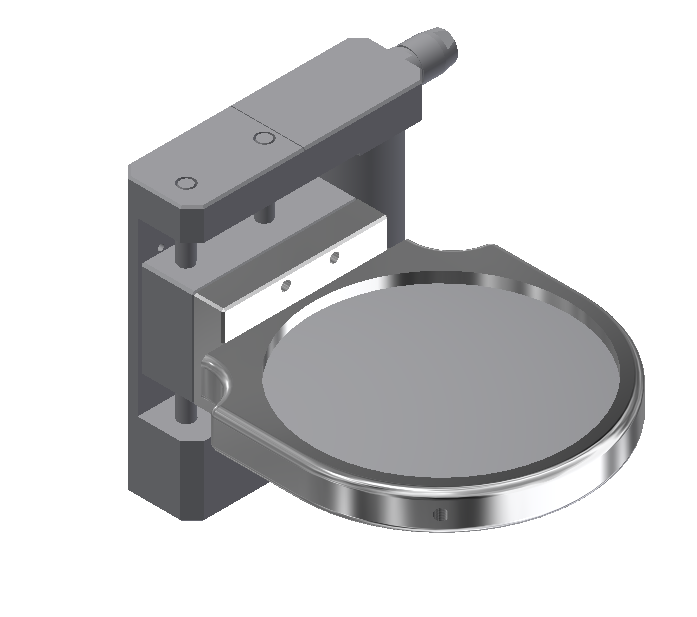}
   \includegraphics[height=5cm]{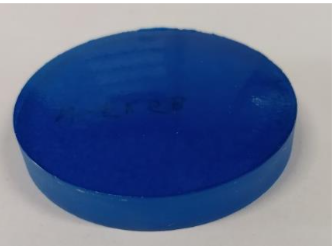}
   \end{tabular}
   \end{center}
   \caption[example] 
   { \label{fig:CAM_CollimatorUnit} 
Left panel: view of the opto-mechanical mount for the collimator lens. Right panel: image of the lens under manufacturing. }
   \end{figure} 
The 65 mm diameter collimator lens is the first element after the CAM selector mirror along the optical path. It will be placed on a linear stage with 15 mm stroke (PI M-111.1DG1) and it will act as a re-focuser. The linear stage is in house and already controlled.
The lens is glued to an Aluminum mount by 3 3M 2216 glue spots (see Figure~\ref{fig:CAM_CollimatorUnit}, left panel).

\subsection{Folding Mirror}
\begin{figure} [h]
   \begin{center}
   \begin{tabular}{c} 
   \includegraphics[height=5cm]{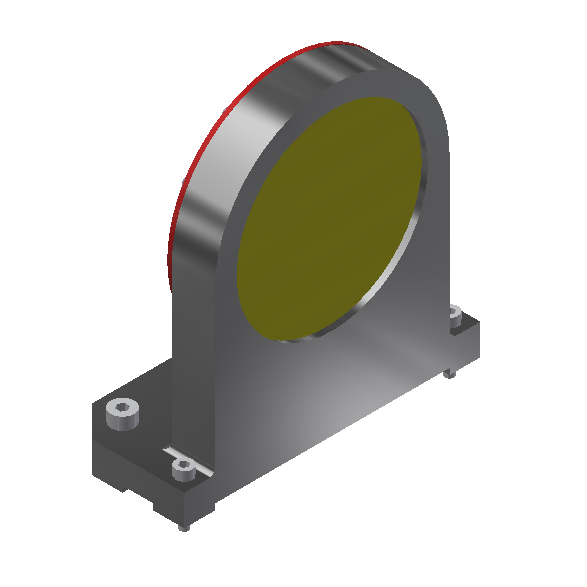}
   \includegraphics[height=5cm]{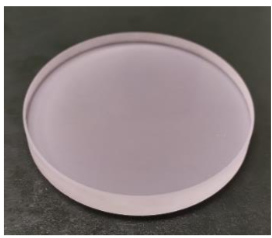}
   \end{tabular}
   \end{center}
   \caption[example] 
   { \label{fig:CAM_FoldingMirrorUnit} 
Left panel: view of the opto-mechanical mount for the folding mirror. Right panel: image of the fused silica part under manufacturing. }
   \end{figure} 
Figure~\ref{fig:CAM_FoldingMirrorUnit} shows the fused silica folding mirror with its opto-mechanical mount (left side), that is at the moment under manufacturing.
The mirror mount will be screwed on the front surface of the external CAM structure. The mirror will be kept in position inside the mount frame by a styrodur ring and a support. The beam is redirected through the folding mirror along an axis parallel to the main telescope axis. 

\subsection{Filter Wheel}
\begin{figure} [h]
   \begin{center}
   \begin{tabular}{c} 
   \includegraphics[height=5.0cm]{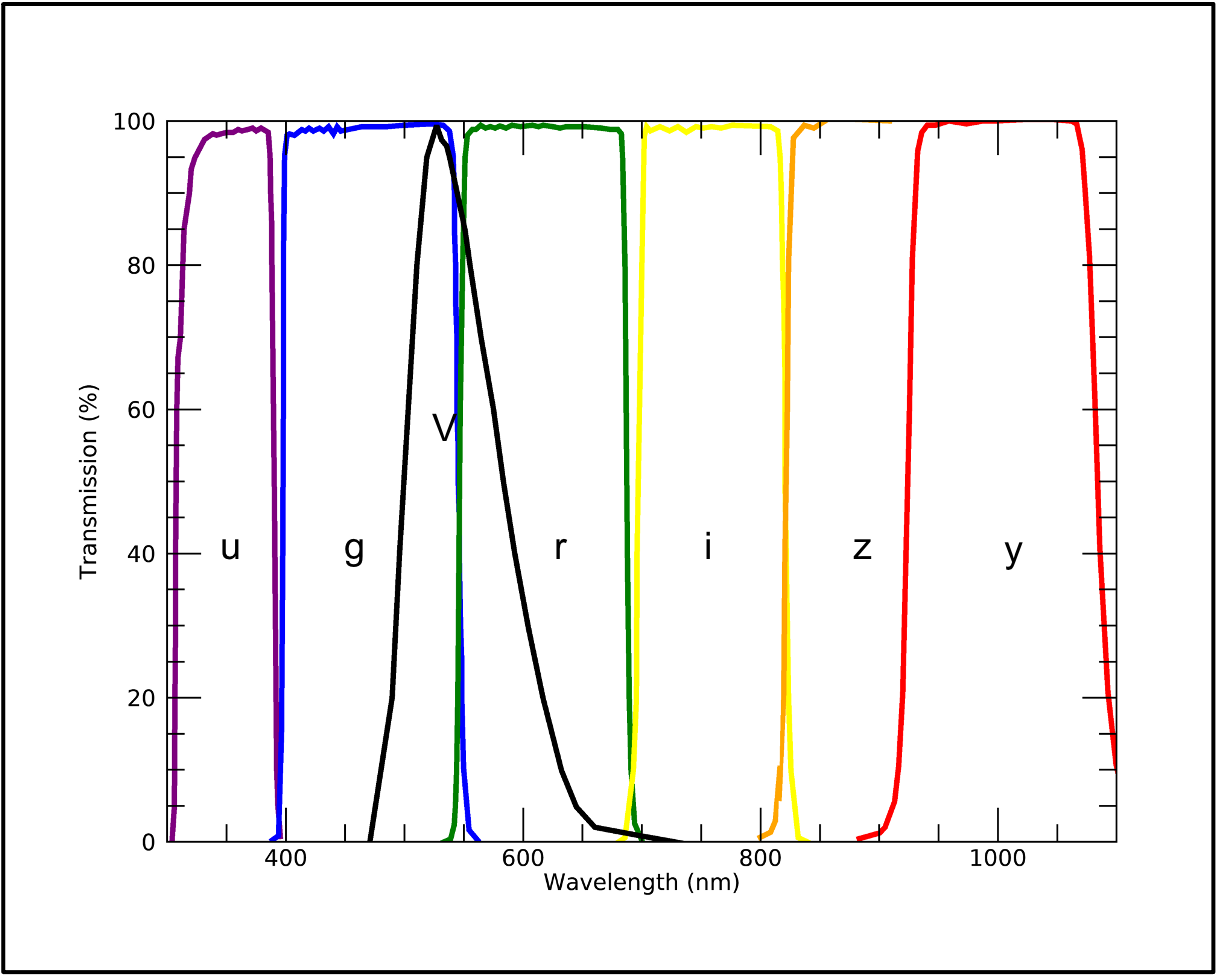}
   \includegraphics[height=5.0cm]{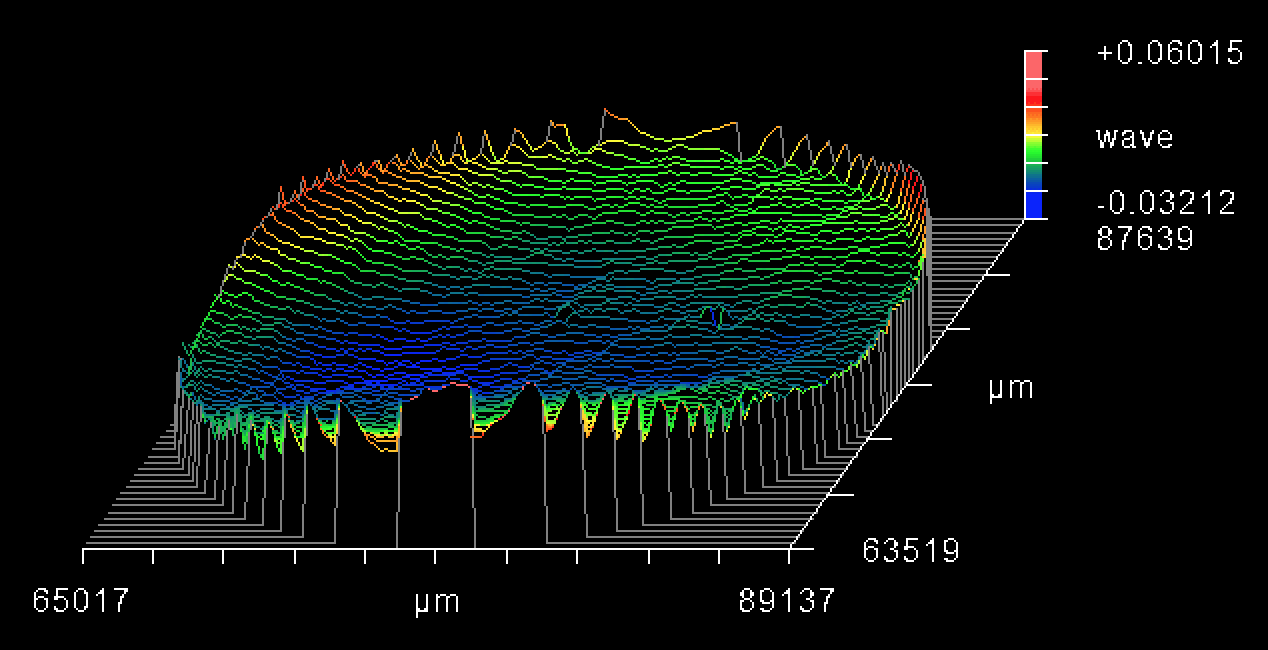}
   \end{tabular}
   \end{center}
   \caption[example] 
   { \label{fig:CAM_filters} 
Left: Transmission of the CAM broad-band filter set as function of wavelength as tested after the production. The filter selection has been optimized to follow the LSST filter passbands. Right: Wavefront Map for the r-band filter. PV = 0.092 wave, rms= 0.011 wave.  }
   \end{figure} 
In front of the camera lens there is a filter wheel made by a rotary stage (PI M‐116.DG) and a custom filter support, which can host up to 8 elements.
A broad-band filter set (\emph{ugrizY} and \emph{V-Johnson}) from Asahi Spectra USA\footnote{https://www.asahi-spectra.com/} are already in house. 
Figure~\ref{fig:CAM_filters}{} shows the filter transmission as function of wavelength as tested after the production.  The filter selection has been optimized to match as best as possible the LSST filter passbands. 
A position of the filter wheel is left empty for the Spectroscopy mode.

\subsection{Camera lens}
The 2 doublets and 2 singlets of max diameter of 30 mm forming the camera lens are under construction. 
The camera lens will relay the Nasmyth focus on the detector, with a F$_{\#}$=3.6. It is planned to include the 2 doublets and the 2 singlets in a tubular structure anchored to the CCD mounting. The optical components will be holded inside the tube by specific spacers and creating 3 3M 2216 glued spots on the side of the element. This operation will be performed via small holes radially drilled inside the support.

\section{CONCLUSIONS}
The development of the CAM system has been seriously affected by the ongoing COVID19 pandemic, interrupting some providers activities and resulting in a delay of about some months. A new optical design has been development, keeping the general design concept, to mitigate the unfeasible waiting time for the procurement of some glasses.
However, all the CAM parts  are  in  an  advanced  phase  of  realization,  very  close  to  be  completed.  The CAM system will be assembled in Italy (INAF-Padua) where also the preliminary integration and final test for the complete instrument SOXS will be done before the shipment to the NTT in La Silla.   The  instrument  integration  and  test  phase  is  planned  in 2021.





\bibliography{report} 
\bibliographystyle{spiebib} 

\end{document}